\documentclass[letter]{aa} % for the letters 
\usepackage{graphicx}
\usepackage{natbib}
\usepackage{epsf}
\usepackage{psfig}
\usepackage{lscape}
\usepackage{txfonts}
\begin{document}
   \title{Detection of diffuse radio emission at large distance from the center of the galaxy cluster A2255}

%   \subtitle{I. Overviewing the $\kappa$-mechanism}

    \author{R.F.Pizzo\inst{1}, A.G. de Bruyn\inst{1,2}, L. Feretti\inst{3}, F. Govoni\inst{4}}

     \institute{Kapteyn Institute, Postbus 800, 9700 AV Groningen, The Netherlands \\ \email{pizzo@astro.rug.nl}
         \and
             ASTRON, Postbus 2, 7990 AA Dwingeloo, The Netherlands \\ 
          \and
             Istituto di radioastronomia-CNR/INAF, via Gobetti 101, 40129 Bologna, Italy\\ 
          \and
             INAF-Osservatorio astronomico di Cagliari, Poggio dei Pini, Strada 54, 09012 Capoterra (CA), Italy\\ }

%   \date{Received September 15, 1996; accepted March 16, 1997}

%\abstract{}{}{}{}{} 
% 5 {} token are mandatory
 
  \abstract{}{Low-frequency radio observations of galaxy clusters are the
    key to detecting the diffuse extended emission associated with
    them. The presence and properties of such radio sources in galaxy clusters reveal the existence of magnetic fields on a large scale and
      allow theories to be tested concerning both the origin of relativistic
      particles in the ICM and their propagation.}{A deep radio observation
    of the A2255 galaxy cluster was carried out at 85 cm with the
    WSRT. The good UV-coverage and sensitivity achieved by these
      observations allowed us to image the complex structure of the low-brightness, extended cluster sources (radio halo and relic).}{ These sources show a larger extent than what has been imaged so far at this frequency, with two new structures located SW and NW of the cluster center and at projected distances of 2 Mpc from it.}{The physical properties of the newly detected structures, together with the active dynamical state of the cluster, support a connection with large-scale structure (LSS) formation shocks.}

   \keywords{galaxies:clusters:general -- galaxies: clusters: individual (Abell 2255) -- galaxies: intergalactic medium -- radio continuum: general}

\authorrunning{Pizzo et al.}
\titlerunning{Detection of new extended emissions in A2255}
   \maketitle
%
%________________________________________________________________

\section{Introduction}

The gravitational hierarchical formation of large-scale structures in the
  universe drives shocks in the intergalactic medium (IGM). These convert the kinetic energy associated with cosmic flows into thermal energy, with IGM temperatures reaching values of 5-10 keV in massive clusters and \textless 1 keV in large-scale filaments \citep[e.g.][]{2002MNRAS.337..199M,2001ApJ...552..473D}.

Shocks can re-accelerate old relativistic electron populations, released  by the former AGN activity within a cluster, or they can also directly accelerate thermal electrons of the IGM.
%It is generally believed that shocks accelerate particles to relativistic
%energies through two possible mechanisms: diffuse shock acceleration of
%thermal electrons of the IGM or adiabatic compression of fossil radio plasma,
%released by an AGN whose central engine has ceased to inject fresh plasma
%\citep[e.g.][]{ensslin,en+krish,2002MNRAS.331.1011E,2007MNRAS.375...77H}. Finding
%relics further away from the center of clusters could discriminate between
%these two possible scenarios.} 
The detection of radio emission from intergalactic shocks has important implications for our understanding of cosmology and astrophysics: it provides a test of structure formation models, can confirm the existence of the undetected warm-hot intergalactic medium, and can trace its distribution \citep{keshet}.

Cluster-wide relativistic electron populations are observed  in several
merging and post-merging clusters as diffuse steep-spectrum structures. They
are grouped in three classes: radio halos, relics, and mini-halos
\citep{fergiov}. Radio halos are unpolarized, extended ($\sim$ 1 Mpc or more) structures located at the cluster center. Relics still have an extended shape, but they lie at the cluster periphery and show high polarization percentages ($\sim$20$\%$). 
%They are found  in clusters both with and without a cooling core, suggesting that they may be related to minor or off-axis mergers, as well as to major mergers. Theoretical models propose that they are tracers of shock waves in merger events (see discussion). 
Mini-halos are detected around a powerful radio galaxy at the center of cooling core clusters and have a typical size of $\sim$ 500 kpc.

These diffuse cluster structures provide important information on the history and evolution of clusters, by improving our knowledge about the presence and importance of both large-scale magnetic fields and relativistic particles in the IGM. Furthermore, since up to now they have only been found in clusters with signatures of merging in the optical and X-ray domains, their detection could  be considered to indicate a perturbed dynamical state. 

The low  radio-surface brightness and steep spectrum make the detection of halos and relics rather difficult. However, in the past few years several works on radio halos and their hosting clusters \citep[e.g.][]{2003tsra.symp..209F,2005xrrc.procE8.02F} have improved our knowledge of these radio sources. On the other hand, the location of relics in the outermost cluster regions makes their detection very problematic because, usually, only the cluster central regions are imaged at radio wavelengths with high sensitivity. Detailed broad band radio studies of relics are still missing; similarly, the comparison of radio with X-ray emission is not possible because of the lack of sensitivity of X-ray satellites in the peripheral cluster regions. A review of the current knowledge of relics is found in \cite{giovferrelics}. One of the main goals of detecting ``relic-like'' structures in galaxy clusters is to constrain their origin.    

The galaxy cluster A2255 is  nearby \citep[z=0.0806,][]{struble} and rich. {\it ROSAT}
X-ray observations indicate that it has recently undergone a merger
\citep{burns,fer,miller}. Recent {\it XMM-Newton} observations of A2255
  revealed temperature asymmetries of the ICM and reached the conclusion that
  the merger happened $\sim$ 0.15 Gyr ago, probably in the E-W
  direction with a still uncertain position angle \citep{sakelliou}. Optical
studies of A2255 reveal kinematical substructures in the form
of several associated groups \citep{yuan}, and the high ratio of velocity
dispersion to X-ray temperature \citep[6.3 KeV;][]{horner} also indicates a
non-relaxed system. When studied at radio wavelengths, A2255 shows a diffuse radio halo (located at the center of the cluster) and a relic source (at the cluster periphery), together with a large number of
embedded head-tail radiogalaxies \citep{har}. This cluster is the first and only one in which polarized radio emission from a radio halo has been
detected.The halo shows filaments of strong polarized emission
  ($\sim$ 20-40 \%) with the magnetic fields fluctuating up to scales of
  $\sim$ 400 kpc in size \citep{gov}. 

We observed A2255 at several radio wavelengths to better understand
the nature of the polarized emission of the filaments in the halo (whether
this is intrinsic or due to a projection effect) and to search for low surface-brightness features located far from the cluster center. We did not observe strong polarization from the halo filaments at 85 cm, but
  the analysis is complicated by the presence of very strong Galactic
  polarization. These polarization results, as well as those at 18, 21, and 25 cm, will
  be discussed in a future paper (Pizzo et al., in prep.). In this Letter we present the recent results of the total intensity imaging of A2255 at 85 cm. 

Throughout this paper we assume a $\Lambda$CDM cosmology with $H_{0}$ =
71 km s$^{-1}$\ Mpc$^{-1}$, $\Omega_{m}=0.3$, and $\Omega_{\Lambda}=0.7$. At
the distance of A2255, 1 \hbox{$\arcsec$} corresponds to 1.5 kpc and $1\hbox{$^{\circ}$} $ = 5.4 Mpc.

%__________________________________________________________________

\section{Observations and reduction}

The observations were conducted with the Westerbork Synthesis Radio Telescope (WSRT)
at 85 cm. The array consists of fourteen 25 m dishes on an east-west baseline and
uses earth rotation to fully synthesize the UV-plane. Ten of the telescopes
 are on fixed mountings, 144 m apart; the four (2 x 2)
remaining dishes are movable along two rail-tracks. In the
array, the baselines can extend from 36 m to 2.7 km.

The 6x12h technique has been used for the observations. For full imaging over the
whole primary beam, six array configurations were used with the four movable
telescopes stepped at 12 m increments (i.e. half the dish diameter) and the shortest
spacing running from 36 m to 96 m. This provides continuous UV-coverage with
interferometer baselines ranging from 36 to 2760 m. The pointing and phase center of
the telescope was directed towards (J2000.0): RA = 17 $^{\rm h}13^{\rm m} 00^{\rm
s}$, Dec = $+64\hbox{$^{\circ}$ }07\hbox{$^{\prime}$ }59\hbox{$\arcsec$}$, which is
located in the middle of A2255.

The feeds/receivers cover the frequency range of about 310-380 MHz. The band
can be completely covered by the new wide band
correlator, which can independently process 8 tunable bands of 10 MHz centered at 315, 324, 332, 341, 350, 359, 367, and 376 MHz. Each band is
covered by 128 channels  in 4 cross-correlations to recover all Stokes
parameters. For our observations, the size of the largest detectable structure is $\sim$ 1\hbox{$^{\circ}$}, the size of the primary beam is $\sim$ 2\hbox{$^{\circ}$}, and the resolution is $\sim$ 1\hbox{$^{\prime}$}.

The data reduction was done with the WSRT-tailored NEWSTAR reduction package
following the standard procedure: Fourier-Transform, Clean, and Restore.
Self-calibration was applied to remove residual phase and gain
variations. The
data were flux-calibrated using 3C295, for which we adopted a flux at 325 MHz
of 64.5 Jy.
Automatic flagging was applied at the beginning to take care of bad data, and
further flagging was done subsequently in an iterative way on the basis of the
selfcal residuals. The total amount of flagged data was $\sim$ 25\% .

The final noise level in the image ranges from about 0.05 mJy/beam to 0.15 mJy/beam
  and is limited by classical confusion noise.

%----------------------------------------------------------------------
\begin{figure*}[th]
\begin{center}
\vspace{7.5 cm}
\includegraphics{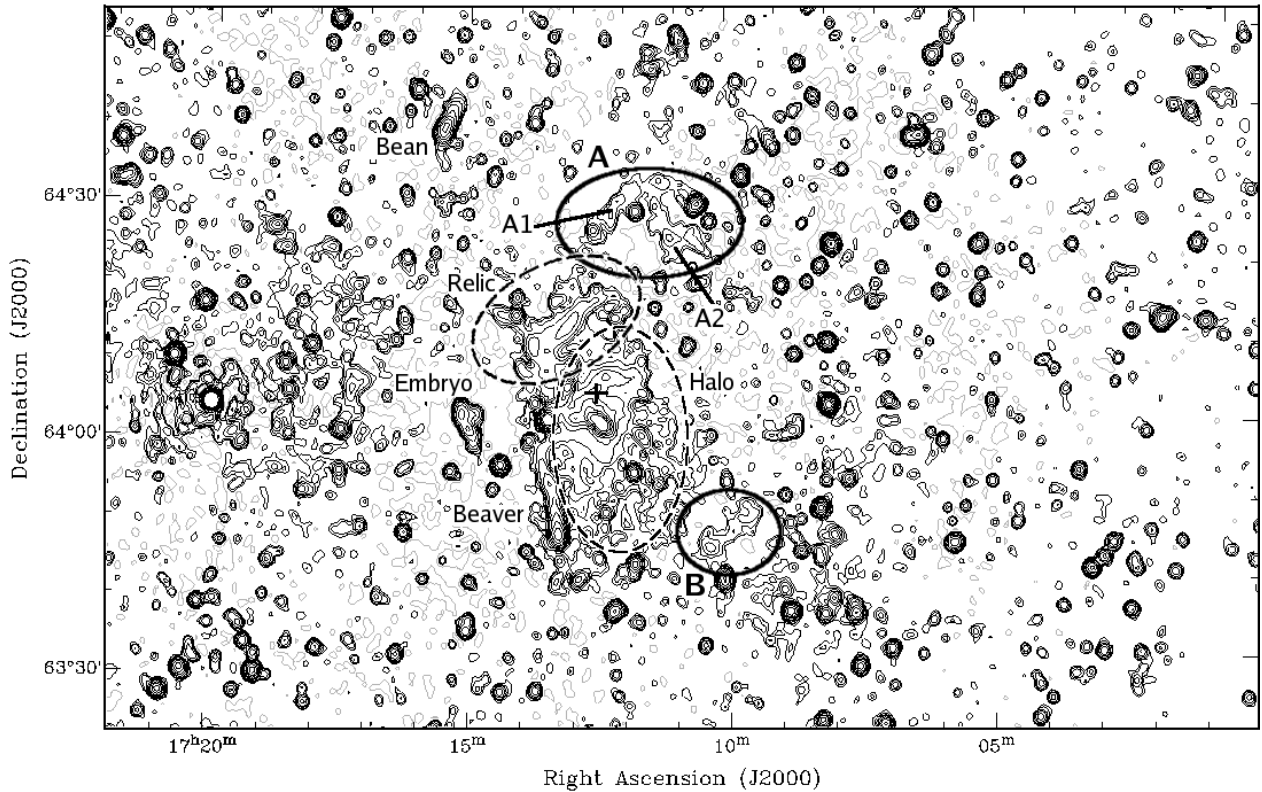}
\vspace{3.5cm}
\caption{Contour map of A2255 at 85 cm. The resolution FWHM is $\sim$~ 1\hbox{$^{\prime}$}. The noise level is $\sim$ 0.1 mJy/beam. The contours are -0.3, 0.3, 0.6, 1.2, 2.4, 4.8, 9.6, 20, 40, 80, 160 mJy/beam. The cross indicates the cluster center. Please note that some of the emission surrounding the bright source 4C +64.21 (RA = 17 $^{\rm h}19^{\rm m} 58^{\rm s}$, Dec = $+64\hbox{$^{\circ}$ }04\hbox{$^{\prime}$ }45\hbox{$\arcsec$}$) is probably due to dynamic range limitations.}
\end{center}
\end{figure*}

%--------------------------------------------------------------------------

\section{Results}

The final 85 cm image of A2255 is shown in Fig. 1. It shows the well known
extended halo (located at the cluster center), the relic (located
10\hbox{$^{\prime}$ } to the northeast from it) and the radio galaxies, at a
very large distance from the center of the cluster: the Embryo (RA = 15 $^{\rm
  h}15^{\rm m} 05^{\rm s}$, Dec = $+64\hbox{$^{\circ}$ }03\hbox{$^{\prime}$
}42\hbox{$\arcsec$}$) and the Beaver (RA = 17 $^{\rm h}13^{\rm m} 19^{\rm s}$,
Dec = $+63\hbox{$^{\circ}$ }48\hbox{$^{\prime}$ }16\hbox{$\arcsec$}$) lie at
$\sim$ 1.6 Mpc from the cluster center, while the Bean (RA = 17 $^{\rm h}15^{\rm m} 31^{\rm s}$, Dec = $+64\hbox{$^{\circ}$ }39\hbox{$^{\prime}$ }28\hbox{$\arcsec$}$) lies at more than 3.5 Mpc \citep[quoted names are taken from][]{har}.

The 85 cm image improves over previous WSRT imaging \citep{fer} by a factor of
20 in sensitivity and reveals a morphology of the extended emission of A2255, that is
much larger than the one known so far. At higher sensitivity, the central
  radio halo looks much more complex than in previous imaging, and it has a more
  extended shape ($\sim$ 500 kpc towards S and SW). Its southern region,
where the Beaver radio galaxy lies, is directly connected with the tail of the
Beaver radio galaxy, which has doubled in length to more than 1 Mpc in the 85
cm image, compared to the 21 cm map \citep{2007..Pizzo..proc}. Moreover, the
already known northern relic shows  a low brightness extent opposite to the
cluster center, which was not visible at 21 cm .\\

  To understand the low-level, very extended emission, we made a low-resolution map (4$^{\prime}$), where the point sources were removed. This image reveals an extended structure, which can be both positive
  and negative, on a scale of 0.5 to 1
  degree, with a T$_{B}$ $\sim$ 0.15 K. We suspect this stems from
  structure in the Galactic foreground. The large-scale negative region that
  we note to the east of A2255 is present in all 8 bands, so we suggest that is also related
  to the large-scale diffuse Galactic structure that is not properly imaged  by the
WSRT, whose shortest spacing is 36 meters.

Two new extended emission regions have now been detected. The new relics (we refer to them in this way) are one order of magnitude
brighter than the ``Galactic'' structures discussed above. They are located NW (labeled A) and SW (labeled B) of the cluster center and lie at a projected distance of 2 Mpc from it.  Previous 21 cm imaging of the cluster \citep{2007..Pizzo..proc} revealed that these sources are genuine features and not a collection of discrete sources.

Another extended emission region seems to be present at low level at
$\sim$ 500 kpc westward of the new  SW relic ( $\sim$ RA = 17 $^{\rm h}08^{\rm m} 45^{\rm s}$, Dec = $+63\hbox{$^{\circ}$ }43\hbox{$^{\prime}$ }45\hbox{$\arcsec$}$). Moreover, a significant, extended feature is present also at the very eastern edge of the cluster. Its shape might suggest a physical association (jet structure) with the radio source WN B1717+6404 (RA = 17 $^{\rm h}17^{\rm m} 28^{\rm s}$, Dec = $+64\hbox{$^{\circ}$ }01\hbox{$^{\prime}$ }17\hbox{$\arcsec$}$) . Our new follow-up observations at the frequency range of 115-165 MHz will tell us more about the nature of this extended feature.\\ 

%There seems to be a large scale negative structure to the east of Abell 2255. It is present in all 8 bands. We do not believe it is due to selfcalibration or cleaning. Instead, we suggest it is related to large scale diffuse Galactic structure that is not properly imaged by the WSRT, which has a shortest spacings of 36 meters.

The newly detected distant relics have different shapes. The SW relic appears like a filament about 8 $\hbox{$^{\prime}$ }$ in length and 2 $\hbox{$^{\prime}$ }$ in width. It has the same orientation of the known NE relic, but is located on the opposite side from the cluster center and at double the distance from it. The NW relic has a more complex morphology. We can distinguish 2  filaments (labels A1 and A2): A1 points towards the cluster center, and has a length of about 8 $\hbox{$^{\prime}$ }$ and a width of about 1 $\hbox{$^{\prime}$ }$; A2 is $\sim$ 13 $\hbox{$^{\prime}$}$ in length and $\sim$ 1 $\hbox{$^{\prime}$}$ in width and is perpendicular to A1. The SW relic  has a total flux density of  $\sim$ 17 mJy, the NW relic of $\sim$ 61 mJy.

We made low-resolution maps of the cluster at 320 and 380 MHz and produced a spectral index map. The estimated values of the spectral index for
the new relics are between -2.5 \textless $\alpha$ \textless -0.5, with
  the steeper values in the regions closer to the cluster center. The typical
error $\sigma_{\alpha}$ $\sim$ $\pm$ 1. A wider spectral range is needed to better determine the radio spectrum of these new features.

%------------------------------------------------------------------------

%\begin{figure*}
%\begin{center}
%\vspace{3.2 cm}
%\special{psfile=NWzoom.ps voffset=110 hoffset=35 vscale=22 hscale=22 angle=270}
%\special{psfile=NWspectralindex.ps voffset=-23 hoffset=500 vscale=25 hscale=25, angle=90}
%\vspace{0.5 cm}
%\caption{Left side: contour map of the NW relic.  The contours are -0.3, 0.3, 0.6, 1.2, 2.4, 4.8, 9.6, 20, 40, 80, 160 mJy/beam. The resolution FWHM is $\sim$~ 1\hbox{$^{\prime}$}. Right side: spectral index map of the NW relic. The resolution FWHM is $\sim$~ 2\hbox{$^{\prime}$}.}.
%\end{center}
%\end{figure*}

%----------------------------------------------------------------------

\section{Discussion}

Relics are found  at the periphery of clusters both with and without a cooling core, suggesting that they may be related to minor or off-axis mergers, as well as to major mergers. The formation of relics is suggested to be related to shocks either by Fermi-I diffuse acceleration of ICM electrons \citep{ensslin,keshet} or  by adiabatic energization of the relativistic electrons confined in fossil radio plasma (``ghosts''), released by a former active radio galaxy \citep{en+krish}. Shocks can be induced by the mergers of sub clusters \citep[``merger shocks'',][]{2001ApJ...561..621R}, or  they can be due to the accretion flows of IGM near the virial radius of the cluster \citep[``accretion shocks'' or ``LSS shocks'',][]{2003MNRAS.342.1009M,2003ApJ...585..128K}. 

Relic radio sources that can be associated with radio ghosts have been found
in several clusters and a spectral index is also available for a few of them. The radio relic in Abell 85 \citep{slee} may be considered a
prototype for this kind of objects, due to its size, morphology, and strong polarization.

 If relics are directly produced by shocks, they should be found farther from the center of  the clusters, where the X-ray surface brightness is very low. The ones  associated with merger shocks should come in pairs and be located on the opposite side of the cluster center along the axis of the merger, with a ring-like structure elongated perpendicular to this direction \citep{2001ApJ...561..621R}. This scenario is observed at radio wavelengths in the galaxy clusters A3376 and A3367, where a pair of large, optically unidentified, diffuse radio sources have been detected \citep{bagchi2006,1997MNRAS.290..577R}. This could also be the case for the previously known NE and the newly  detected SW relics in A2255, located on opposite sides with respect to the cluster center (although not symmetrically) and showing the same orientation, perpendicular to an NE-SW axis, that could describe the recently undergone merger.
% The asymmetry might be intrinsic and due to a larger velocity of the shock in the southern part of the cluster than in the northern one, probably caused by inhomogeneous IGM. Alternatively, the asymmetry can be explained with projection effects. 
The origin of the extended ring-like features of A3376 is probably due to merger
shocks that accelerate the thermal electrons of the hot ICM or old
relativistic particles to relativistic energies. On the other hand, in A2255
the new relics lie at the virial radius of the cluster \citep{neumann} and at
a large projected distance from the X-ray emission associated with the cluster. This makes them more likely associated with shocks waves in the cosmic environment than with a hot ICM. 

%Their symmetric and tangential juxtaposition relative to the merger axis suggests an association with LSS shocks. This seems to be the case also for the previously known NE and the newly  detected SW relics in A2255. They are located at opposite sides with respect to the cluster center (although not symmetrically) and they show the same orientation, perpendicular to the axis describing the former merger. The asymmetry should be intrinsic and due to a larger velocity of the shock in the southern part of the cluster than in the northern one, probably caused by inhomogeneous IGM. Alternatively, the asymmetry can be explained with projection effects. 

We note that the new relics of A2255 are reminiscent of the filamentary structures of the halo detected by \cite{gov}, which are
elongated and organized in a kind of ``net'' or ``web''. This suggests that
these structures could also be related to LSS or merger shocks and would
naturally explain the high percentage of linear polarization. 
%Within this picture they would only appear in projection to be in the cluster center.}
%This is very puzzling and could indicate that we are actually detecting something not strictly confined within the cluster, but on a large scale and related with the LSS.

The spectra of known LSS-related relics are steep ($\alpha$ $\sim$ -1.2), and they are found to be linearly polarized at the level of 10\%-30\%. The previously known NE relic shows such a polarization \citep{gov}, but we cannot confirm the presence of polarized flux for the newly detected structures so far. Using RM synthesis \citep{br2005}, we have generated an RM-cube over a wide range of Faraday depths. It shows that most of the polarized emission in the field of A2255 is due to the Galactic foreground, since it extends well beyond the cluster ``boundary'', and it has no counterpart in total intensity; however, there are several features in the RM-cube suggesting a link with continuum structures belonging to the cluster. Part of the NW relic may have been detected at a Faraday depth of about -28 rad/m$^2$, far from the depth of the bulk of the Galactic polarized emission (Pizzo et al., in prep.).

The spectral indices of the new relics (-2.5 \textless $\alpha$ \textless -0.5 with $\sigma_{\alpha} \pm$ 1) seem to agree with the typical value of LSS shocks, although we should not expect a spectrum steeper than $\alpha \sim -1.5$ for LSS-related structures, since electrons are constantly accelerated at the shock front (priv. comm. with Matthias Hoeft). As the shock dissipates, the relativistic particles age, explaining the higher values of $\alpha$ we detect. Moreover, the trend shown by the spectral index, which is flatter at the periphery of the cluster and steepening towards the cluster center, is what one also expects for shock-related structures \citep[see the relic sources in A3667, ][]{roettiger}. However, a wider spectral range is needed to confirm this result.

The morphology of the new structures, together with their size ($\simeq$ 1 Mpc), location (tangential with respect to the cluster center and at the virial radius of the cluster), spectra, and (possible) polarization support an LSS shock origin.

\section{Conclusions}
 
We have presented the results of new 85 cm WSRT observations of the galaxy cluster A2255.

\begin{itemize}
 
\item Our map reveals that the extended radio emission in A2255 is both much more extended and complex than known before \citep{fer}.

\item Two new extended emission regions are detected at a projected distance of 2 Mpc from the cluster center.

\item The spectral index of the new structures ranges between -2.5 \textless $\alpha$ \textless -0.5, with $\sigma_{\alpha} \pm$ 1. 

\item The morphology of the new structures, together with their size ($\simeq$ 1 Mpc), location (tangential with respect to the cluster center and at the virial radius of the cluster), spectra, and (possible) polarization support an LSS shock origin.

\end{itemize}

In the current models of large-scale structure formation, matter is
distributed in filaments and sheets that are all connected. Radio emission is
expected to be present in collapsing filaments, where the efficiency of
electrons acceleration is higher than in cluster centers \citep{bagchi2002}. We are now reaching the sensitivity needed for detecting radio emission from regions in between galaxy clusters. This is suggested by the discovery of elongated diffuse emission (0917+75) at $\sim$ 3.8 Mpc from the nearest rich cluster \citep{dewdney,harris1993,2000NewA....5..335G}, of a relic source in A2069 \citep{1999NewA....4..141G} at 4.6 Mpc from the cluster center and of extended diffuse radio emission coincident with the filament of galaxies ZwCl2341.1+0000, 2.5 Mpc in size \citep{bagchi2002} .
The steep spectrum of the diffuse  inter-galactic radio sources and,
  based on current theoretical results \citep[e.g.][]{2002MNRAS.337..199M},
  their connection with shocks in the IGM predict, at low radio frequencies, a sizable
  population of as yet undetected extended emissions around galaxy clusters. A recent {\it ROSAT}  X-ray survey observation indicates that A2255 belongs to the north ecliptic pole super cluster \citep{mullis}. Therefore, it is possible that there are signs of interaction between A2255 and several other galaxy clusters. A panoramic study of A2255 out to many degrees ($\sim$ 10 Mpc) is needed to search for such interaction. The detection of new filaments and relics related to  LSS shocks will be enormously improved through observations with the next generation of low-frequency radio telescopes like LOFAR, LWA, and SKA, for which we are preparing ourselves.

\begin{acknowledgements}
 The Westerbork Synthesis Radio Telescope is operated by ASTRON (Netherlands Foundation for Research in Astronomy) with support from the Netherlands Foundation for Scientific Research (NWO).
\end{acknowledgements}

\bibliographystyle{aa}
\bibliography{9304}

\begin{thebibliography}{34}
\expandafter\ifx\csname natexlab\endcsname\relax\def\natexlab#1{#1}\fi

\bibitem[{{Bagchi} {et~al.}(2006){Bagchi}, {Durret}, {Neto}, \&
  {Paul}}]{bagchi2006}
{Bagchi}, J., {Durret}, F., {Neto}, G.~B.~L., \& {Paul}, S. 2006, Science, 314,
  791

\bibitem[{{Bagchi} {et~al.}(2002){Bagchi}, {En{\ss}lin}, {Miniati}, {Stalin},
  {Singh}, {Raychaudhury}, \& {Humeshkar}}]{bagchi2002}
{Bagchi}, J., {En{\ss}lin}, T.~A., {Miniati}, F., {et~al.} 2002, New Astronomy,
  7, 249

\bibitem[{{Brentjens} \& {de Bruyn}(2005)}]{br2005}
{Brentjens}, M.~A. \& {de Bruyn}, A.~G. 2005, \aap, 441, 1217

\bibitem[{{Burns}(1998)}]{burns}
{Burns}, J.~O. 1998, Science, 280, 400

\bibitem[{{Dav{\'e}} {et~al.}(2001){Dav{\'e}}, {Cen}, {Ostriker}, {Bryan},
  {Hernquist}, {Katz}, {Weinberg}, {Norman}, \& {O'Shea}}]{2001ApJ...552..473D}
{Dav{\'e}}, R., {Cen}, R., {Ostriker}, J.~P., {et~al.} 2001, \apj, 552, 473

\bibitem[{{Davis} {et~al.}(2003){Davis}, {Miller}, \& {Mushotzky}}]{miller}
{Davis}, D.~S., {Miller}, N.~A., \& {Mushotzky}, R.~F. 2003, \apj, 597, 202

\bibitem[{{Dewdney} {et~al.}(1991){Dewdney}, {Costain}, {McHardy}, {Willis},
  {Harris}, \& {Stern}}]{dewdney}
{Dewdney}, P.~E., {Costain}, C.~H., {McHardy}, I., {et~al.} 1991, \apjs, 76,
  1055

\bibitem[{{En{\ss}lin} {et~al.}(1998){En{\ss}lin}, {Biermann}, {Klein}, \&
  {Kohle}}]{ensslin}
{En{\ss}lin}, T.~A., {Biermann}, P.~L., {Klein}, U., \& {Kohle}, S. 1998, \aap,
  332, 395

\bibitem[{{En{\ss}lin} \& {Gopal-Krishna}(2001)}]{en+krish}
{En{\ss}lin}, T.~A. \& {Gopal-Krishna}. 2001, \aap, 366, 26

\bibitem[{{Feretti}(2003)}]{2003tsra.symp..209F}
{Feretti}, L. 2003, in Texas in Tuscany. XXI Symposium on Relativistic
  Astrophysics, ed. R.~{Bandiera}, R.~{Maiolino}, \& F.~{Mannucci}, 209--220

\bibitem[{{Feretti}(2005)}]{2005xrrc.procE8.02F}
{Feretti}, L. 2005, in X-Ray and Radio Connections (eds. L.O. Sjouwerman and
  K.K Dyer) Published electronically by NRAO,
  http://www.aoc.nrao.edu/events/xraydio Held 3-6 February 2004 in Santa Fe,
  New Mexico, USA, (E8.02) 10 pages, ed. L.~O. {Sjouwerman} \& K.~K. {Dyer}

\bibitem[{{Feretti} {et~al.}(1997){Feretti}, {Boehringer}, {Giovannini}, \&
  {Neumann}}]{fer}
{Feretti}, L., {Boehringer}, H., {Giovannini}, G., \& {Neumann}, D. 1997, \aap,
  317, 432

\bibitem[{{Feretti} \& {Giovannini}(1996)}]{fergiov}
{Feretti}, L. \& {Giovannini}, G. 1996, in IAU Symposium, Vol. 175,
  Extragalactic Radio Sources, ed. R.~D. {Ekers}, C.~{Fanti}, \&
  L.~{Padrielli}, 333--+

\bibitem[{{Giovannini} \& {Feretti}(2000)}]{2000NewA....5..335G}
{Giovannini}, G. \& {Feretti}, L. 2000, New Astronomy, 5, 335

\bibitem[{{Giovannini} \& {Feretti}(2004)}]{giovferrelics}
{Giovannini}, G. \& {Feretti}, L. 2004, Journal of Korean Astronomical Society,
  37, 323

\bibitem[{{Giovannini} {et~al.}(1999){Giovannini}, {Tordi}, \&
  {Feretti}}]{1999NewA....4..141G}
{Giovannini}, G., {Tordi}, M., \& {Feretti}, L. 1999, New Astronomy, 4, 141

\bibitem[{{Govoni} {et~al.}(2005){Govoni}, {Murgia}, {Feretti}, {Giovannini},
  {Dallacasa}, \& {Taylor}}]{gov}
{Govoni}, F., {Murgia}, M., {Feretti}, L., {et~al.} 2005, \aap, 430, L5

\bibitem[{{Harris} {et~al.}(1980){Harris}, {Lari}, {Vallee}, \& {Wilson}}]{har}
{Harris}, D.~E., {Lari}, C., {Vallee}, J.~P., \& {Wilson}, A.~S. 1980, \aaps,
  42, 319

\bibitem[{{Harris} {et~al.}(1993){Harris}, {Stern}, {Willis}, \&
  {Dewdney}}]{harris1993}
{Harris}, D.~E., {Stern}, C.~P., {Willis}, A.~G., \& {Dewdney}, P.~E. 1993,
  \aj, 105, 769

\bibitem[{{Horner}(2001)}]{horner}
{Horner}, D. 2001, {PhD thesis}

\bibitem[{{Keshet} {et~al.}(2004){Keshet}, {Waxman}, \& {Loeb}}]{keshet}
{Keshet}, U., {Waxman}, E., \& {Loeb}, A. 2004, New Astronomy Review, 48, 1119

\bibitem[{{Keshet} {et~al.}(2003){Keshet}, {Waxman}, {Loeb}, {Springel}, \&
  {Hernquist}}]{2003ApJ...585..128K}
{Keshet}, U., {Waxman}, E., {Loeb}, A., {Springel}, V., \& {Hernquist}, L.
  2003, \apj, 585, 128

\bibitem[{{Miniati}(2002)}]{2002MNRAS.337..199M}
{Miniati}, F. 2002, \mnras, 337, 199

\bibitem[{{Miniati}(2003)}]{2003MNRAS.342.1009M}
{Miniati}, F. 2003, \mnras, 342, 1009

\bibitem[{{Mullis} {et~al.}(2001){Mullis}, {Henry}, {Gioia}, {B{\"o}hringer},
  {Briel}, {Voges}, \& {Huchra}}]{mullis}
{Mullis}, C.~R., {Henry}, J.~P., {Gioia}, I.~M., {et~al.} 2001, \apjl, 553,
  L115

\bibitem[{{Neumann}(2005)}]{neumann}
{Neumann}, D.~M. 2005, \aap, 439, 465

\bibitem[{{Pizzo} \& {de Bruyn}(2007)}]{2007..Pizzo..proc}
{Pizzo}, R.~F. \& {de Bruyn}, G. 2007, in Extragalactic Jets: Theory and
  Observation from Radio to Gamma Ray, ed. T.~A. {Rector} \& D.~S. {De Young}

\bibitem[{{Ricker} \& {Sarazin}(2001)}]{2001ApJ...561..621R}
{Ricker}, P.~M. \& {Sarazin}, C.~L. 2001, \apj, 561, 621

\bibitem[{{Roettiger} {et~al.}(1999){Roettiger}, {Burns}, \&
  {Stone}}]{roettiger}
{Roettiger}, K., {Burns}, J.~O., \& {Stone}, J.~M. 1999, \apj, 518, 603

\bibitem[{{Rottgering} {et~al.}(1997){Rottgering}, {Wieringa}, {Hunstead}, \&
  {Ekers}}]{1997MNRAS.290..577R}
{Rottgering}, H.~J.~A., {Wieringa}, M.~H., {Hunstead}, R.~W., \& {Ekers}, R.~D.
  1997, \mnras, 290, 577

\bibitem[{{Sakelliou} \& {Ponman}(2006)}]{sakelliou}
{Sakelliou}, I. \& {Ponman}, T.~J. 2006, \mnras, 367, 1409

\bibitem[{{Slee} {et~al.}(2001){Slee}, {Roy}, {Murgia}, {Andernach}, \&
  {Ehle}}]{slee}
{Slee}, O.~B., {Roy}, A.~L., {Murgia}, M., {Andernach}, H., \& {Ehle}, M. 2001,
  \aj, 122, 1172

\bibitem[{{Struble} \& {Rood}(1999)}]{struble}
{Struble}, M.~F. \& {Rood}, H.~J. 1999, \apjs, 125, 35

\bibitem[{{Yuan} {et~al.}(2003){Yuan}, {Zhou}, \& {Jiang}}]{yuan}
{Yuan}, Q., {Zhou}, X., \& {Jiang}, Z. 2003, VizieR Online Data Catalog, 214,
  90053

\end{thebibliography}

\end{document}